\documentclass[aps,prb,reprint,amsmath,superscriptaddress,amssymb,graphicx]{revtex4-1}
\usepackage{amsfonts}
\usepackage{bbm}
\usepackage{graphicx}
\usepackage{dcolumn}
\usepackage{bm}
\usepackage{subfigure}
\usepackage{amssymb}
\newcommand{\me}{\mathrm{e}} \newcommand{\mi}{\mathrm{i}}\newcommand{\piup}{\mathrm{\pi}}
\newcommand{\dif}{\mathop{}\!\mathrm{d}}
\DeclareMathAlphabet{\mathfsl}{OT1}{cmss}{m}{sl}
\begin{document}
	\title{Phonon-Assisted Andreev Reflection at Majorana Zero Mode}
	\author{Ning Dai}	
	\affiliation{International Center for Quantum Materials, School of Physics, Peking University, Beijing 100871, China}
\author{Qing-Feng Sun}
\email[]{sunqf@pku.edu.cn}
\affiliation{International Center for Quantum Materials, School of Physics, Peking University, Beijing 100871, China}
\affiliation{Beijing Academy of Quantum Information Sciences,
West Bld.\#3,No.10 Xibeiwang East Rd., Haidian District, Beijing 100193,China}
\affiliation{Collaborative Innovation Center of Quantum Matter, Beijing 100871, China}
	\date{\today}

\begin{abstract}
One of the typical features of Majorana zero mode (MZM) at the end of topological superconductor
is a zero-bias peak in the tunneling spectroscopy of the normal lead-superconductor junction.
In this paper we study on a model with one phonon mode coupling to the superconductor lead of
the normal lead-superconductor junction, which can be viewed
as an electron-lead/phonon-coupled-MZM/hole-lead structure.
The phonon-coupled MZM acts as a series of channels in which electron can turn into hole
by absorbing and emitting phonons.
These channels present in the local density of states (LDOS) as a series of stripes,
generating the corresponding peaks in the tunneling spectroscopy.
In LDOS, the electron-phonon interaction narrows and redistributes the weight among stripes.
In the tunneling spectroscopy, the heights of peaks present a feature of the multi-phonon process.
With these investigations, our work illuminates the mechanism of phonon-assisted Andreev reflection
at a Majorana zero mode.
\end{abstract}
\maketitle

\section{introduction}

Majorana fermion, a fermion which is its own antiparticle,
can be achieved in condensed matter as a quasiparticle of electrons and holes.\cite{MF1,MF2,MF4,MF5,MFadd}
For example, Majorana fermion appears in spinless p wave superconductor
at the transition between strong and weak pairing phases.\cite{ps1,ps2,ps3,ps4}
In practice, the Majorana zero mode (MZM) can be achieved at the end of a quantum wire
which is in proximity to a s-wave superconductor and in a Zeeman field.\cite{MZM1,MZM2}
An efficient method to check the existence of MZM is to construct
a normal lead-superconductor junction and measure the tunneling spectroscopy.
With a MZM residing in the junction, the tunneling spectroscopy presents
a zero-bias peak\cite{ZBP1,ZBP2,YSZheng1,WP2}
which is resulted from the Majorana induced resonance Andreev reflection.\cite{MIRAR}
Up to now, the zero-bias peak has been observed in many experiments,\cite{ZBPep9,ZBPep5,ZBPep2,ZBPep4,ZBPep6,ZBPep7,ZBPep8}
but the height of the peak usually is much lower than the perfect value $2 e^2/h$,
because of the soft gap,\cite{soft} thermal effect,\cite{thermal1,thermal2}
or the electron-phonon interaction (EPI).
Very recent, an experiment work has successfully achieved the zero-bias peak with
the peak height $ 2e^2/h$
in an indium antimonide semiconductor nanowire covered a superconducting shell,\cite{ZBPep10}
which gives a strong evidence of the existence of the MZM.

The phonon plays an important role in electron transport.\cite{PA1,PA3,PA4,PA5,SunPhonon1,SJT1}
While short wave phonon participates in the superconducting pairing,\cite{psc1,psc2}
it is also worth to consider the influence of long wave optical phonon
in the Majorana induced Andreev reflection.
Some reports have shown the spectroscopy of phonon-assisted Andreev reflection,
where a series of peaks appear with the equal peak interval $\frac{\omega_p}{2}$
with $\omega_p$ being the frequency of the phonon.\cite{PAAR2,Gramich1,FTF6,add1,add2}
However, the Majorana induced Andreev reflection is a kind of special Andreev reflection
which is more like resonance tunneling.
The normal lead plays a role of both electron lead and hole lead,
and the MZM acts as a quantum dot with a single zero energy level.\cite{MIRAR}
In phonon-assisted resonance tunneling, the local density of states (LDOS) of the quantum dot
and the differential conductance present a series of peaks at $n\omega_p$ with the integer $n$
and are sensitive to the average Fermi energy of the two leads,\cite{PART1,PART2}
which is exactly zero for electron and hole leads in the case of
Majorana induced Andreev reflection and aligns to the MZM.
Moreover, very recent, Song and Das Sarma have studied the relaxation time of the Majorana qubit
with the coupling of phonons, and show that the relaxation rate can be manipulated by
the accompanied phonon energy which can be tuned by the voltage bias.\cite{addSong}
So far, the mechanism of phonon-assisted Majorana induced Andreev reflection
has not been investigated, which is exactly the subject of the present work.

In order to look into the influence of phonon on Majorana induced Andreev reflection,
we investigate a simplified model with one phonon mode coupling to
the superconductor of a normal lead-superconductor junction in this paper.
An non-equilibrium Green's function formalism is developed to numerically solve
the LDOS and conductance of this normal lead-superconductor junction.
With the phonon mode coupling, MZM generates a series of channels
in which electron can turn into hole by absorbing and emitting phonons.
These channels present in the LDOS as a series of stripes
which have the same width in energy, the same shape in real space but different scales.
The stripes are reduced to peaks in energy distribution at a fixed position.
The coupling strength of the normal lead and superconductor
has a broadening effect on the peaks, but has no influence on the weight of each peak.
On the other hand, the EPI strength restrains the broadening
and elevates the weight of non-zeroth phonon sidebands.
The heights of peaks grow with the EPI strength moderately at zero temperature
but intensively at nonzero temperature.
However, the soar of peaks in LDOS with temperature is wiped out
in conductance by thermal broadening effect.
At zero temperature, the width of peaks in conductance are exactly equal to their counterpart in LDOS,
and the height of the $n$th conductance peak presents a feature of $n$-phonons process.

The rest of this paper is organized as follows:
In Sec.\ref{sec2} we introduce the normal lead-superconductor model
and the non-equilibrium Green's function,
of which the detailed deduction is developed in the Appendix.
The results are presented in Sec.\ref{result}.
Specifically, we first give a picture of phonon assisted Andreev reflection
at MZM in Sec.\ref{sec31}, then numerically study the LDOS in Sec.\ref{spfu}
and differential conductance in Sec.\ref{cond}, respectively.
In the end, we summary the work in Sec.\ref{sec4}.

\section{\label{sec2}model and method}

\begin{figure}
	\includegraphics[width=\linewidth]{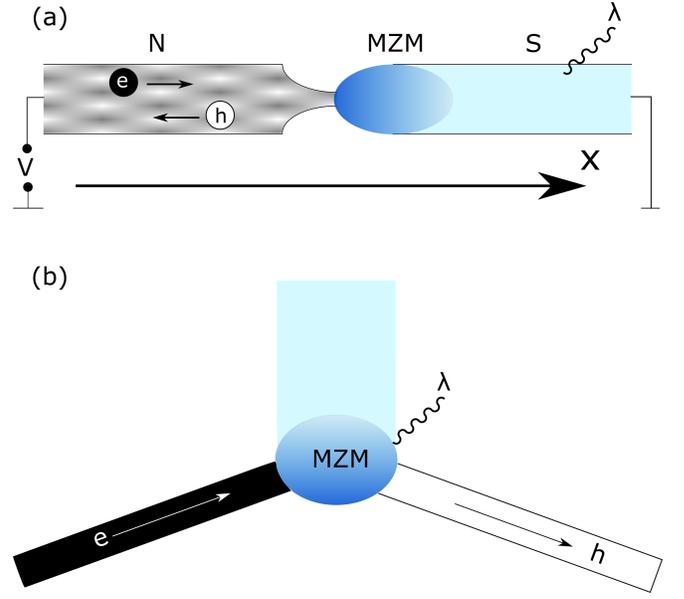}
	\caption{\label{device}
(a) Andreev reflection in weak tunneling 1D normal lead-superconductor junction
with a phonon mode coupled to the superconductor wire.
A MZM exists at the end of the superconductor wire.
The bias $V$ is added on the normal lead and the superconductor is grounded.
(b) Regarding the normal wire as a hybrid of an electron lead and a hole lead,
this normal lead-superconductor junction can be viewed
as an electron-lead/phonon-coupled-MZM/hole-lead structure.}
\end{figure}

The phonon-coupled normal lead-superconductor device is shown in Fig.\ref{device}(a).
With the help of Zeeman field and proximity to s-wave superconductor,
an 1D quantum wire changes into a spinless p-wave superconductor
or a Kitaev chain\cite{Kitaev} of which a MZM resides at the end.
In this paper, a single phonon mode is coupled to this superconductor
to simulate the effect of long wave optical phonon,
of which the frequency is $\omega_p$ and the momentum can be omitted.
The experimental realization of such a model can be achieved with
the help of some material with special phonon spectrum and the spin-orbit coupling,
e.g. a nanowire with spin-orbit coupling and the vibration phonon mode.
In fact, it has been experimentally found that
phonon modes exist in various systems, such as nanowires, nanotubes,
quantum dots and so on.
For example, if the 1D quantum wire is made of carbon nanotube,
there exists a radial breathing mode corresponding to the global radial vibration.\cite{RBM1,RBM2,RBM3}
In addition, the intrinsic vibration phonon mode can exist in nanowires, nanotubes, quantum dots
and some molecules.\cite{Gramich1,addphn1,addphn2,addphn3}

The phonon-coupled superconducting wire is connected to a normal lead,
forming a normal lead-superconductor junction [Fig.\ref{device}(a)].
The superconducting lead is grounded and a bias $V$ is added on the normal lead.
The EPI strength is $\lambda$, and in this paper we also use
the dimensionless quantity $g=\lambda^2/\omega_p^2$ to denote the strength of EPI.
The tunneling strength between the normal lead and superconductor is labeled by $1/\alpha$,
which means a larger $\alpha$ represents a weaker tunneling. The parameter $\alpha$ is called reversed coupling strength in the rest of this paper.
The Hamiltonian of this device is presented below:
\begin{eqnarray}
H&=&H_L+H_R+H_T ,\label{1}\\
H_R&=&H_{R0}+\lambda(\hat{a}+\hat{a}^\dagger)\sum_k\hat{c}_k^\dagger\hat{c}_k
 +\omega_p\hat{a}^\dagger\hat{a}, \label{HR}\\
H_{R0}&=&\sum_k\varepsilon_k\hat{c}_k^\dagger\hat{c}_k+\Delta_k\hat{c}_{k}\hat{c}_{-k}
 +\Delta_k^*\hat{c}_{-k}^\dagger\hat{c}_k^\dagger, \label{HR0}\\
H_L&=&\sum_k\varepsilon'_k\hat{d}^\dagger_k\hat{d}_k, \label{HL} \\
H_T&=&\sum_{k,k'}(T_0/\alpha)\hat{d}^\dagger_k\hat{c}_{k'}+\mathrm{h.c.},\label{HT}
\end{eqnarray}
where $H_L$ and $H_R$ represent the normal and superconductor lead, respectively,
and $H_T$ describe the coupling term or the tunneling term between the normal lead and superconductor.
In particular, $H_R$ consists of spinless p-wave superconductor term $H_{R0}$,
EPI term $\lambda(\hat{a}+\hat{a}^\dagger)\hat{c}_k^\dagger\hat{c}_k$
and phonon term $\omega_p\hat{a}^\dagger\hat{a}$.
The operators $\hat{a}(\hat{a}^\dagger)$, $\hat{c}(\hat{c}^\dagger)$
and $\hat{d}(\hat{d}^\dagger)$ annihilate (create) a phonon,
an electron in normal lead and in superconducting lead, respectively.
The energy spectrum $\varepsilon_k$ and $\varepsilon'_k$ are both parabolic,
and the constant $T_0$ stand for the normal lead-superconductor coupling strength at $\alpha=1$.

The EPI term in Eq.(\ref{HR}) complicates the solving of this system.
A well developed canonical transformation method is widely used
to solve problems in this kind.\cite{SunPhonon1,PART1,PAAR2,po}
Introducing the canonical transformation on any operator $\hat{O}$,
that $\hat{\bar{O}}=\me^S\hat{O}\me^{-S}$ with
\begin{equation}
S=\sum_k\hat{c}_k^\dagger\hat{c}_k\frac{\lambda}{\omega_p}(\hat{a}^\dagger-\hat{a}).
\end{equation}
The creation and annihilation operators under this transformation are:
\begin{eqnarray}
	\hat{\bar{c}}_k&=&\me^S \hat{c}_k \me^{-S}=
\hat{c}_k\me^{-\frac{\lambda}{\omega_p}(\hat{a}^\dagger-\hat{a})},\label{fieldoperator1}\\
\hat{\bar{c}}^\dagger_k
&=&\me^S \hat{c}^\dagger_k \me^{-S}
=\hat{c}_k^\dagger\me^{\frac{\lambda}{\omega_p}(\hat{a}^\dagger-\hat{a})},\label{fieldoperator2}\\
	\hat{\bar{a}} &=&\me^S \hat{a} \me^{-S} =
\hat{a}-\frac{\lambda}{\omega_p}\sum_k\hat{c}_k^\dagger\hat{c}_k ,\label{fieldoperator3}\\
\hat{\bar{a}}^\dagger
&=&\me^S \hat{a}^\dagger \me^{-S}
=\hat{a}^\dagger-\frac{\lambda}{\omega_p}\sum_k\hat{c}_k^\dagger\hat{c}_k ,\label{fieldoperator4}
\end{eqnarray}
and the operators $\hat{d}_k$ and $\hat{d}^{\dagger}_k$ remain unchange under this transformation.

Before applying this canonical transformation,
we note that the superconducting pairing term in Eq.(\ref{HR0}) originates from
the attraction interaction $U_{k,k'}$ between electrons,
that is, the original form of Hamiltonian $H_{R0}$ in Eq.(\ref{HR0}) is
\begin{equation}
H_{R0}=\sum_k\varepsilon_k\hat{c}_k^\dagger\hat{c}_k
+\sum_{k,k'}U_{k,k'}\hat{c}^\dagger_{k'}\hat{c}^\dagger_{-k'}\hat{c}_{-k}\hat{c}_{k} .\label{HR0'}
\end{equation}
A mean-field approximation is applied on Eq.(\ref{HR0'})
to arrive at the BCS Hamiltonian [Eq.(\ref{HR0})], where
\begin{equation}
	\Delta_k\equiv\sum_{k'}U_{k,k'}\langle\hat{c}^\dagger_{-k'}\hat{c}^\dagger_{k'}\rangle.\label{superpair}
\end{equation}
The canonical transformation should act on Eq.(\ref{HR0'}) rather than Eq.(\ref{HR0}).
Under the canonical transformation,
i.e. substituting Eqs.(\ref{fieldoperator1}-\ref{fieldoperator4}) to
Eqs.(\ref{HR},\ref{HL},\ref{HT},\ref{HR0'}),
the Hamiltonians $H_{R0}$ and $H_L$ remain unchange with
$\bar{H}_L =\me^S H_L \me^{-S} = H_L$ and $\bar{H}_{R0}= \me^S H_{R0} \me^{-S} = H_{R0}$.
On the other hand, the Hamiltonians $H_{R}$ and $H_T$ change into:
\begin{eqnarray}
\bar{H}_R&=& \bar{H}_{R0}
-\frac{\lambda^2}{\omega_p}\sum_{k,k'}\hat{c}^\dagger_k\hat{c}^\dagger_{k'}\hat{c}_{k'}\hat{c}_k
-\frac{\lambda^2}{\omega_p}\sum_{k}\hat{c}^\dagger_k\hat{c}_k
+\omega_pa^\dagger a , \label{mean2}\\
\bar{H}_T&=&
\sum_{k,k'}(T_0/\alpha)\me^{-\frac{\lambda}{\omega_p}(\hat{a}^\dagger-\hat{a})}\hat{d}^\dagger_k\hat{c}_{k'}
+\mathrm{h.c.}.\label{8}
\end{eqnarray}
By using the mean-field approximation same to what we have used
in arriving at BCS Hamiltonian, Eq.(\ref{mean2}) is reduced to
\begin{equation}
	\bar{H}_R=\sum_k\bar{\varepsilon}_k\hat{c}^\dagger_k\hat{c}_k+\omega_pa^\dagger a+\sum_k\bar{\Delta}_k\hat{c}_k\hat{c}_{-k}+\mathrm{h.c.},\label{9}
\end{equation}
where
\begin{eqnarray}
\bar{\Delta}_k &=& \Delta_k
+\frac{\lambda^2}{\omega_p}\langle \hat{c}^\dagger_k\hat{c}^\dagger_{-k}\rangle,\\
\bar{\varepsilon}_k &=&\varepsilon_k- {\lambda^2}/{\omega_p}.
\end{eqnarray}
So, the transformed Hamiltonian $\bar{H} =\me^S H\me^{-S} $ reads
\begin{equation}
\bar{H}= H_L+\bar{H}_T+\bar{H}_R.
\end{equation}
In $\bar{H}$, instead of the complicated EPI term $\lambda(\hat{a}+\hat{a}^\dagger)\hat{c}_k^\dagger\hat{c}_k$,
EPI only exists in $\bar{H}_T$ and can be eliminated by a mean-field approximation\cite{SunPhonon1,PART1,PAAR2}
\begin{equation}
\me^{-\frac{\lambda}{\omega_p}(\hat{a}^\dagger-\hat{a})}=\langle\me^{-\frac{\lambda}{\omega_p}(\hat{a}^\dagger-\hat{a})}\rangle=\me^{-g(N+1/2)},\label{appro}
\end{equation}
where $g=\lambda^2/\omega_p^2$ and
\begin{equation}
N=\frac{1}{\me^{\omega_p/k_B \mathcal{T}}-1}\label{N}
\end{equation}
stands for the number of phonon with the temperature $\mathcal{T}$.
The approximation in Eq.(\ref{appro}) includes all the irreducible self-energy of single
normal lead-superconductor coupling ones and single EPI ones,
only omits the high order of irreducible Feynman diagram
which contain both normal lead-superconductor coupling and EPI.
Eq.(\ref{appro}) is exact at $T_0/\alpha=0$ or $\lambda=0$,
and it is reasonable while $T_0/\alpha$ or $\lambda$ is the smallest energy scales,
i.e $T_0/\alpha \ll \min(\lambda,\Delta)$ or $\lambda \ll \min(T_0/\alpha,\Delta)$.\cite{SunPhonon1,PART1,PAAR2}
In addition, at the low temperature, the phonon number $N$ is small
and the effective EPI is weak. So this approximation is
more valid at the low temperature than at high temperature.

The Hamiltonian $\bar{H}$ is discretized to adapt to numerical calculation.
Let $\varepsilon'_k=\frac{\hbar^2 k^2}{2m}-\epsilon_L$,
$\bar{\varepsilon}_k=\frac{\hbar^2 k^2}{2m}-\epsilon_R$, $T_0=\frac{-1}{2ma_0^2}$,
and the order parameter for p wave pairing could be set to $\bar{\Delta}_k=\Delta k$.
These terms with $k$ and $k^2$ can be discretized by the following relation:
\begin{eqnarray}
\sum_kkc_k^\dagger c_k&=&\sum_x\frac{-\mi}{2a_0}[c^\dagger_xc_{x+1}-c^\dagger_xc_{x-1}],\label{de1}
\\
\sum_kk^2c_k^\dagger c_k&=&\sum_x\frac{-1}{a_0^2}[c^\dagger_xc_{x+1}-2c^\dagger_xc_x+c^\dagger_xc_{x-1}].\label{de2}
\end{eqnarray}
In addition, by using the inverse Fourier transformation
$\hat{d}_x=\sum_{k}\me^{-\mi kx}\hat{d}_k$ and $\hat{c}_x=\sum_{k}\me^{-\mi kx}\hat{c}_k$,
we have $\hat{d}_0=\sum_{k} \hat{d}_k$ and $\hat{c}_0=\sum_{k}\hat{c}_k$.
To combine Eqs.(\ref{de1},\ref{de2}) with the inverse Fourier transformation,
the $\bar{H}$ can be discretized into the following form:
\begin{eqnarray}
H_L&=&(\frac{1}{ma_0^2}-\epsilon_L)\hat{d}^\dagger_x\hat{d}_x-\frac{1}{2ma_0^2}\hat{d}^\dagger_{x+1}\hat{d}_x+\mathrm{h.c.}\nonumber\\
&&\hspace{6cm}x\leq0\label{dis1}\\
\bar{H}_R&=&(\frac{1}{ma_0^2}-\epsilon_R)\hat{c}^\dagger_x\hat{c}_x-\frac{1}{2ma_0^2}\hat{c}^\dagger_{x+1}\hat{c}_x+\frac{\Delta}{2a_0}\hat{c}_{x+1}\hat{c}_x+\mathrm{h.c.}\nonumber\\
&&\hspace{6cm}x\geq0\label{dis2}\\
\bar{H}_T&=&-\frac{1}{\alpha}\frac{1}{2 ma_0^2}\me^{-g(N+1/2)}\hat{c}^\dagger_0\hat{d}_0+\mathrm{h.c.}.\label{dis3}
\end{eqnarray}
Here $a_0$ denotes the discretization length.
The normal lead-superconductor interface is located at $x=0$.

\begin{figure}
	\includegraphics[width=\linewidth]{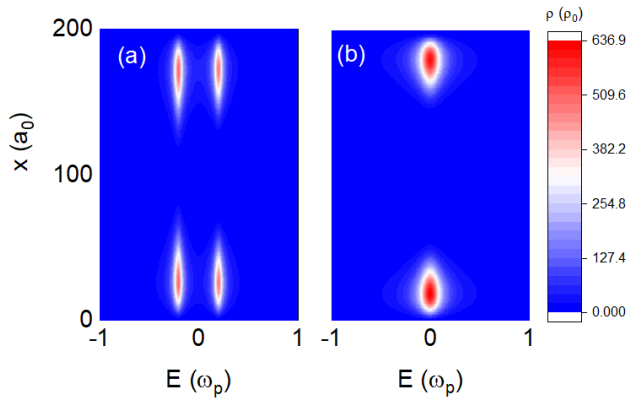}
	\caption{\label{add}
The LDOS of finite p-wave superconductor of different length,
coupling with normal conductor leads in both ends.
The superconductor is discretized into 200 lattices in both panels (a) and (b),
but the lattice lengths are different in (a) shorter system with $a_0=0.05$
and (b) longer system with $a_0=0.07$. }
\end{figure}

With Eq.(\ref{dis1}-\ref{dis3}) we can easily acquire the transformed Green's function $\bar{\mathcal{G}}(x,x')$.\cite{Datta,ZhangYT1,YFZ}
With the help of $\bar{\mathcal{G}}(x,x')$, we are able to achieve the
LDOS $\rho(x,E)$ of phonon-coupled MZM as well as the tunneling spectroscopy $G(V)$.
The explicit formulation is developed in the Appendix of this paper.
The numerical investigations in LDOS and transporting property are detailed in Sec.\ref{result},
which are at the parameters of $m=0.025$, $a_0=0.05$, $\epsilon_R=15$,
$\Delta=20$, $\epsilon_L=2000$, where the natural unit is adopted and
the phonon frequency $\omega_p=1$ as the energy unit.
At this group of parameters the superconducting gap is $E_\mathrm{gap}=28$.

Eqs.(\ref{dis1}-\ref{dis3}) describe the semi-infinite p-wave superconducting wire
coupled with a normal conductor lead,
which has topological edge states in the interface of the superconductor and normal conductor.
If for a finite p-wave superconducting wire
(or a finite p-wave superconducting wire coupled with two normal conductor leads, i.e.
the normal conductor/finite p-wave superconducting wire/normal conductor system),
there exist the Majorana zero modes in both end of the wire.
Fig.\ref{add} shows the LDOS of finite p-wave superconducting wire
coupled with two normal conductors for the different length.
In a short superconducting wire, the MZMs in both ends of the wire
couple with each other, splitting in energy spectrum [see Fig.\ref{add}(a)].
On the other hand, for the long superconducting wire,
the coupling between both ends vanishes,
leaving an isolated MZM in each end of the wire [see Fig.\ref{add}(b)].
This clearly indicates the existence of the MZM in the interface of
the superconductor and normal conductor.

\section{\label{result}result}
\subsection{\label{sec31}mechanism}

\begin{figure}
	\includegraphics[width=\linewidth]{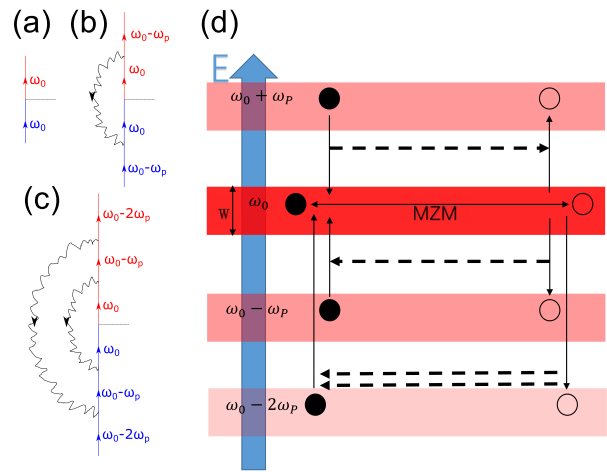}
	\caption{\label{model}
(a-c) Diagrams of Andreev reflections concerning with 0, 1, 2 phonons, respectively.
The wavy line represents the phonon propagator,
and the blue and red lines represent electrons and holes respectively.
(d) The MZM is expanded in energy by normal lead-superconductor coupling
and plays a role of electron-hole exchange channel.
There exist a series of subchannels which are integer number of phonon frequencies apart from the MZM.
Electrons in subchannel can pass through the MZM channel by absorbing and emitting phonons.
The transparency of each channel is expressed by the color.}
\end{figure}

\begin{figure}
	\includegraphics[width=\linewidth]{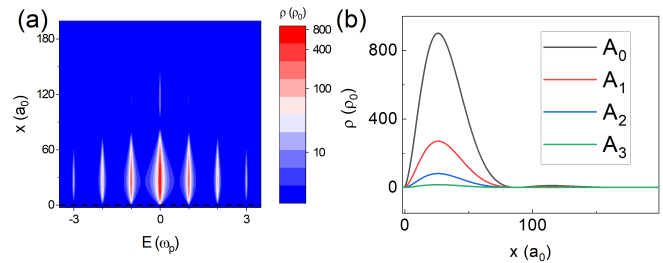}
	\caption{\label{fig1}
(a) The LDOS $\rho(x,E)$ presents several stripes, referring to the channels in Fig.\ref{model}(d).
Logarithmic color bar is applied to exaggerate the existence of narrow stripes.
(b) The space distribution of the stripes.
The curve $A_n$ is extracted from (a) along the track of $E=n\omega_p$.
The shape of $A_n$s are exactly the same but vary in scale.
The normal lead-superconductor interface is at x=0.}
\end{figure}

In condition of weak tunneling and small bias comparing to the superconducting gap,
the normal current through the normal lead-superconductor junction is blocked,
and the current flow is dominated by superconducting current which comes from Andreev reflection at the MZM.
In this case the normal lead can be viewed as a combination of
an electron lead and a hole lead and the normal lead-superconductor junction
can be viewed as a two terminal system [Fig.\ref{device}(b)].
The superconductor lead is suspended and
the remaining electron-lead/MZM/hole-lead structure constitutes a resonance tunneling system
(i.e., the Majorana induced resonance Andreev reflection)
in which the MZM plays the role of the quantum dot.\cite{QD1}
The electron and hole leads are coupled to the MZM with the same tunneling strength,
leading to a perfect tunneling at $V=0$, i.e., the zero-bias peak in the tunneling spectroscopy.
Here the MZM acts as an electron-hole channel at zero energy.
This e-h channel is broadened in energy by the coupling of the normal lead to MZM,
which presents in the tunneling spectroscopy as the widening of the zero-bias peak.
Figure.\ref{model}(a) gives the diagram of the Majorana induced resonance Andreev reflection.
While $\omega_0=0$ is in the e-h channel of MZM,
Fig.\ref{model}(a) has a non-vanishing contribution, leading to an Andreev reflection.

Electrons outside the MZM channel are unable to participate an Andreev reflection
unless with the help of phonons.
The single phonon process is presented in Fig.\ref{model}(b).
If the energy of electron is $\omega_p$ away from $\omega_0=0$,
this electron can also participate the process in Fig.\ref{model}(a) by emitting a phonon.
The multi phonon processes are similar,
that an electron whose energy is $n\omega_p$ away from $\omega_0$ can go through the process
in Fig\ref{model}(a) by emitting $n$ phonons.
Figure.\ref{model}(c) gives an example of $n=2$ case.

\begin{figure*}
	\includegraphics[width=\textwidth]{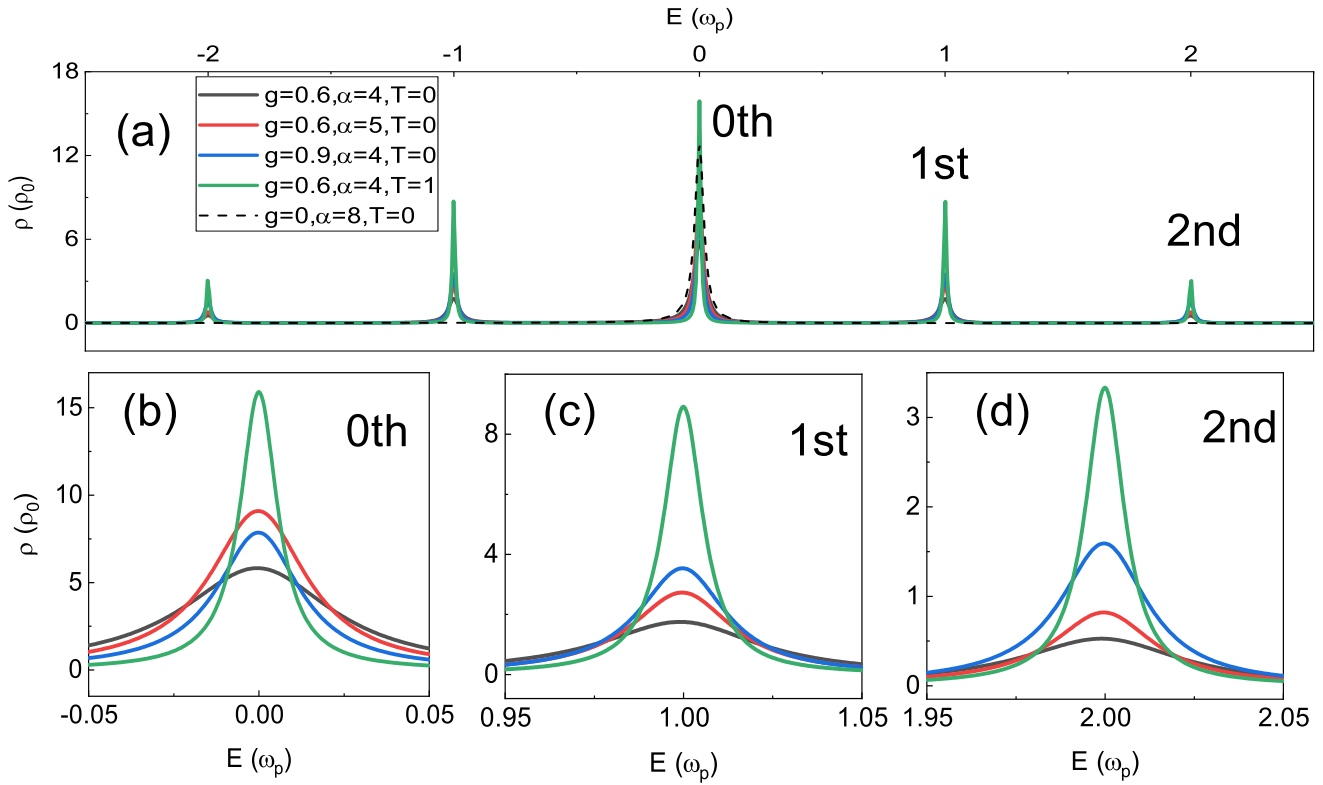}
	\caption{\label{fig2}
(a) The LDOS $\rho(E)$ near the normal lead-superconductor interface
under different groups of parameters. The $n$th phonon peak appears at $E=n\omega_p$.
(b-d) Zoom in on the 0th, 1st, 2nd phonon sideband for nonzero $g$, respectively.
The peaks are narrowed with the increase of EPI strength $g$,
reversed coupling strength $\alpha$ and temperature $\mathcal{T}$.
The area of non-zeroth peaks also increase with $g$
(blue line comparing to black line) and $\mathcal{T}$ (green line comparing to black line).
When EPI is absent, the LDOS is reduced to a single peak at $E=0$ (dashed line).}
\end{figure*}

The no-phonon, single-phonon, and multi-phonon processes in Fig.\ref{model}(a-c)
are summarized in Fig.\ref{model}(d),
which gives a understandable picture of phonon-assisted Andreev reflection at MZM.
The phonon-coupled MZM is a series of e-h channels,
and the $n$th channel locates at $E=n\hbar\omega_p$.
The 0th channel is just the broadened MZM which is perfectly transparent.
The non-zeroth channels, on the other hand, are the copies of the 0th channel.
An electron in the $n$th channel has a chance to take road of the 0th channel
by emitting or absorbing $n$ phonons.
For this reason all channels have the same width $W$ originating from the 0th channel.
In the case of weak EPI, the multi phonon process is infrequent
so that the high order channels are nearly blocked.

\subsection{\label{spfu}Local Density of States}

The channels in Fig.\ref{model}(d) can be directly recognized in the LDOS $\rho(x,E)$,
which is defined by Eq.(\ref{defA}) in the Appendix A.
The LDOS $\rho(x,E)$ gives the probability of the electron at position $x$ having the energy of $E$.
In Fig.\ref{fig1}(a) we numerically solve the $\rho(x,E)$ with $g=0.6$, $\alpha=4$, $\mathcal{T}=0$
in units of $\rho_0=\frac{a_0\omega_p}{10000}$.
The bias voltage $V$ which exists in Eq.(\ref{Aexp}) in
form of $\mu$ ($\mu=eV$) actually has no effect on $\rho$,
because the average Fermi level of electron-lead and hole-lead is
always 0 at every $V$ and exactly aligns to the MZM.
Since the bulk gap $E_\mathrm{gap}\gg E$ in the concerned region
and the tunneling is weak enough to suppress the influence of bulk states near the MZM,
the electron-hole symmetry is conserved,
and $\rho(x,E)$ presents a series of narrow stripes near the normal lead-superconductor interface.
The $n$th stripe is at the position of $E=n\omega_p$,
which is related to an n-phonons process at the MZM.
The total integral of these stripes are $\frac{1}{2}$,
which is corresponding to one Majorana fermion, and is equivalent to half a Fermion.\cite{Kitaev}
Because of the negligible momentum, the long wave phonon splits MZM in energy spectrum
but has little influence in the space distribution,
i.e., the space distribution are the same for every stripes,
which are shown in Fig.\ref{fig1}(b).
The stripes have the same profile but decline in scale with the raise of $n$,
because in the condition of weak EPI the multi phonon process is hard to achieve.

\begin{figure}
	\includegraphics[width=\linewidth]{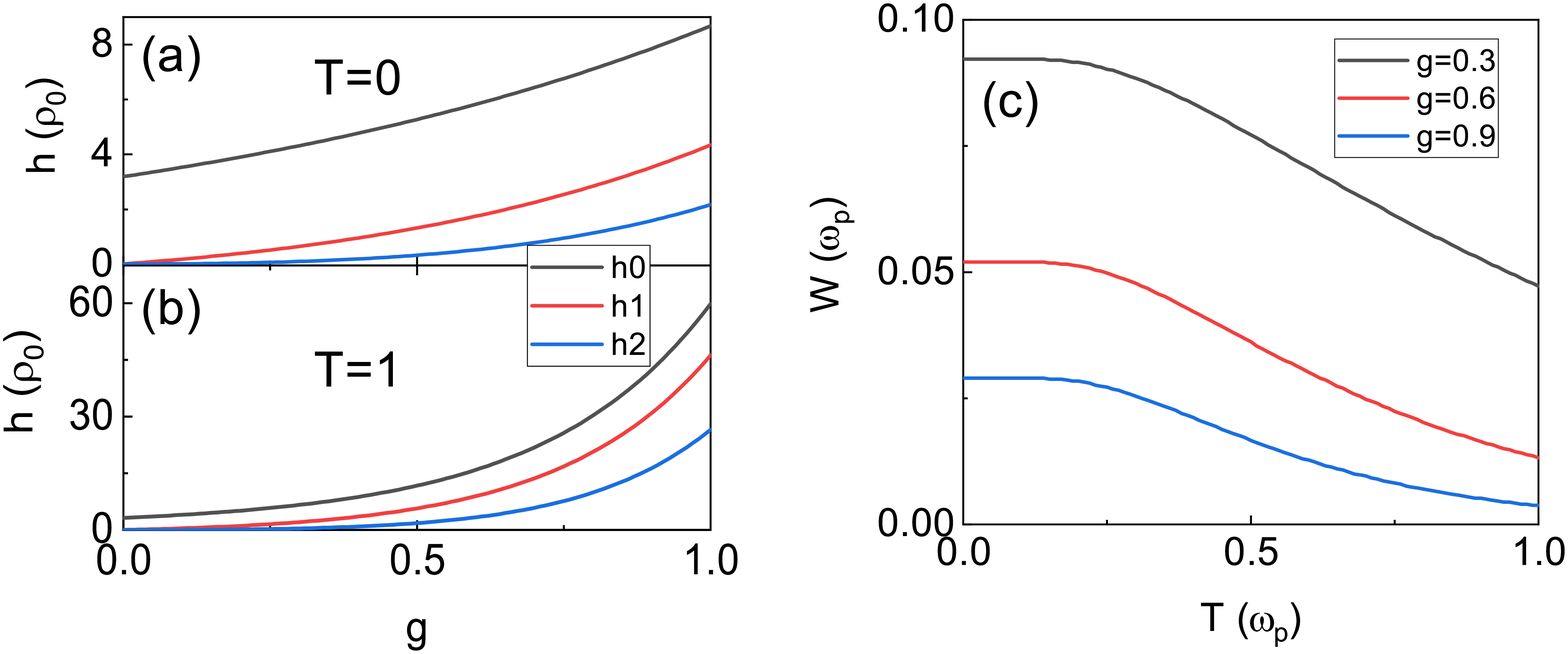}
	\caption{\label{fig3}
The height of the $n$th LDOS peak $h_n$ varies with
the EPI strength $g$ at the temperature $\mathcal{T}=0$ (a) and $\mathcal{T}=1$ (b).
At $g=0$ only $h_0$ is nonzero.
For every $n$, $h_n$ increase with $g$, but the increasement at $\mathcal{T}=1$
is much intenser than $\mathcal{T}=0$.
Moreover, the ratio $h_1/h_0$ and $h_2/h_0$ are larger at $\mathcal{T}=1$ than $\mathcal{T}=0$.
(c) The width of MZM peak varies with temperature $\mathcal{T}$ at different $g$.
These curves have similar profile with the phonon distribution [Eq.(\ref{N})].}
\end{figure}

\begin{figure*}
	\includegraphics[width=\textwidth]{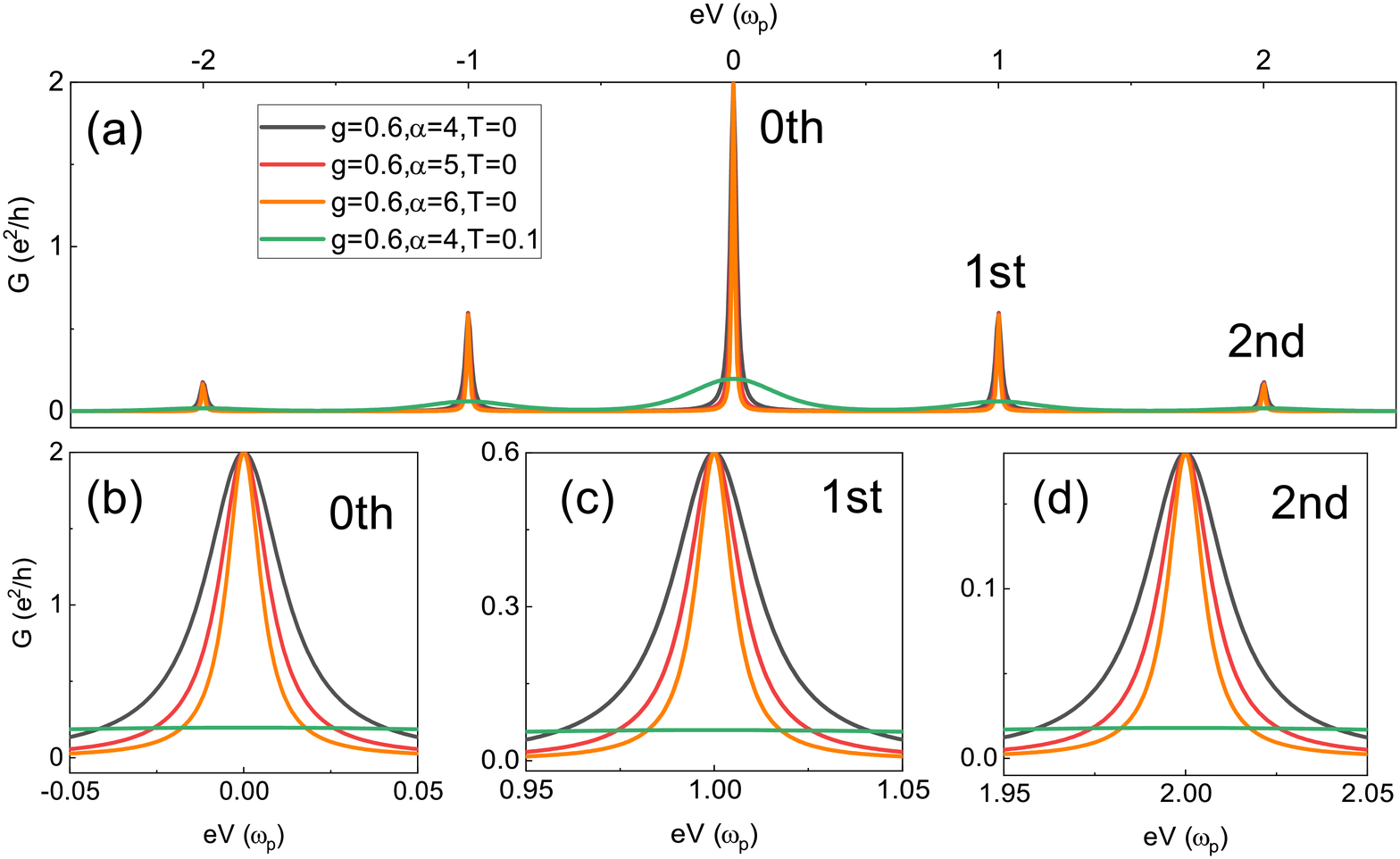}
	\caption{\label{fig4}
(a) The conductance $G$ versus bias voltage $V$ under different groups of parameters.
The $n$th phonon peak appears at $eV=n\omega_p$.
(b-d) Zoom in on the 0th, 1st, 2nd phonon sideband, respectively.
The width of peaks in $G(V)$ is same to peaks in $\rho(E)$ within the same parameter
at $\mathcal{T}=0$, but severely broadened and strongly reduced at $\mathcal{T}=1$.
At $\mathcal{T}=0$ the heights of peaks are unrelated to the normal lead-superconductor
coupling strength $1/\alpha$.}
\end{figure*}

The transport property is dependent on the energy distribution $\rho(E)$
near the normal lead-superconductor interface, which is shown in Fig.\ref{fig2}.
However, Fig.\ref{fig1}(b) indicates that $\rho(E)\equiv0$ at the exact $x=0$ position,
so the average $\rho(E)$ over the small lattice $a_0$ next to the interface,
$\rho(E) =\frac{1}{a_0} \int_0^{a_0} \rho(x,E)dx$, is shown in Fig.\ref{fig2}.
Since the space distributions for every stripe are in the same shape,
Fig.\ref{fig2} honestly reflects the energy distribution near the interface.
The overall outline of $\rho(E)$ is shown in Fig.\ref{fig2}(a),
in which the cross section of stripes in Fig.\ref{fig1}(a) are turned into peaks.
The peaks in $\rho(E)$ represent the electron-hole channels
by which electrons participate Andreev reflection and generate superconducting current.
Figure.\ref{fig2}(b-d) zoom in on the 0th-2nd peaks.
The LDOS is influenced by normal lead-superconductor coupling strength $1/\alpha$,
EPI factor $g$ and temperature $\mathcal{T}$.
The coupling strength $1/\alpha$ directly broadens the MZM in energy
from a delta-type function to a Lorentzian curve,
so it directly affects the width of each peak.
The EPI factor $g$ and temperature $\mathcal{T}$ suppress
the effective normal lead-superconductor coupling in $\bar{H}_T$
by a suppression factor $\me^{-g(N+1/2)}$ and narrow the peaks.
However, the EPI also appears in the coefficient $L_n$[Eq.(\ref{Ln})],
which redistributes the weights of each phonon peak.
So $g$ and $\mathcal{T}$ change the areas of each peak but $\alpha$ does not.
With the decrease of EPI, the LDOS is centralizing in the main peak.
When the EPI is absent, all phonon sidebands are absorbed in the central peak.

The influence of EPI on LDOS is studied in Fig.\ref{fig3}.
The height of LDOS peak $h_n\equiv \rho(n\omega_p)$ describes
the possibility of electron at the energy of $n\omega_p$,
corresponding to a $n$-phonon process.
At $g=0$ where EPI is absent, $h_n=0$ for all nonzero $n$.
From Fig.\ref{fig3}(a) and (b), we can see that all $h_n$s raise with
the EPI strength $g$ for two different reasons:
the suppression in effective coupling strength
between the normal lead and superconductor (see $\bar{H}_T$)
and the redistribution among the peaks.
The increase of $h_0$ is because of the suppression of the effective coupling strength,
which narrow the peaks.
On the other hand for nonzero $n$, the raise of the peak height $h_n$
is due to not only the suppression of the effective coupling strength
but also the inter peak redistribution.
The increase of $h_n$ with $g$ is much intenser at the temperature $\mathcal{T}=1$ than $\mathcal{T}=0$.
The primary reason for this difference is about the coupling strength suppression,
because $h_0$ also soars at $\mathcal{T}=1$ which is irrelevant to the inter peak redistribution.
The suppression factor $\me^{-g(N+1/2)}$ is much smaller at finite temperature
since the phonon number $N$ is positive comparing to $N=0$ at zero temperature.
However, the inter peak redistribution also differs between the temperature $\mathcal{T}=0$ and $\mathcal{T}=1$.
It is obvious that the ratio $h_1/h_0$ and $h_2/h_0$ raise faster at $\mathcal{T}=1$ than $\mathcal{T}=0$,
which mean the single-phonon and bi-phonon processes happen more frequently at the higher temperature.
This is reasonable since at zero temperature electron can only emit phonons and transmit phonons to itself,
while at finite temperature electron can also absorb phonons from the environment.
While the height of peaks is under the co-effect
of the suppression of effective coupling strength and inter peak redistribution,
the width of peaks directly reflects the effective coupling strength [see Fig.\ref{fig3}(c)].
The widths vary slowly at the low temperature region and then begin to drop.
These $W$-$\mathcal{T}$ curves have a similar profile with the Boson distribution of phonons [Eq.(\ref{N})],
because the influence of temperature on width just appears
in the effective coupling strength at $\bar{H}_T$ in form of $N$.

\subsection{\label{cond}conductance}
\begin{figure}
	\includegraphics[width=\linewidth]{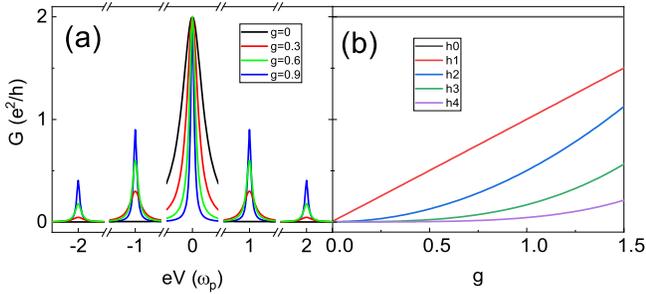}
	\caption{\label{fig5}
(a) Tunneling spectroscopy at different EPI strength $g$ with $\alpha=4$ and $\mathcal{T}=0$.
(b) The heights of conductance peaks vary with $g$ at zero temperature.
The height of $n$th conductance peak $h_n$ is proportional to $g^n$.}
\end{figure}

In Fig.\ref{fig4}, we investigate the effects of the reversed coupling strength $\alpha$
and temperature $\mathcal{T}$ on the tunneling spectroscopy.
The differential conductance $G(V)=\frac{\dif I}{\dif V}$ presents a series of peaks
at $V=n\omega_p$ which just resembles the LDOS.
The resemblance also exist in the effect of the reversed coupling strength $\alpha$,
that the peaks narrow with the increase of $\alpha$.
In fact, at zero temperature the width of peaks in $G(V)$ is exactly the same
with its counterpart in $\rho(E)$,
since the Fermi distribution for electrons is reduced to a step function $\theta(E-V)$.
The energy distribution is uniform away from $E=V$,
so that the electron number in a channel varies only if the bias $V$ is varying in this channel.
On the other hand, peaks in $G(V)$ are severely broadened
and the heights of peaks are strongly reduced at finite temperature,
while the channel itself is slightly narrowed [see Fig.\ref{fig2}].
This is because of the thermal broadening effect.
The thermal broadening is much stronger than the EPI-induced channel narrowing,
which leads to the different thermal behavior of $\rho(E)$ and $G(V)$.

Another difference between $\rho(E)$ and $G(V)$ is
that the heights of peaks vary with the reversed coupling strength $\alpha$ in $\rho(E)$ but do not in $G(V)$.
While normal lead-superconductor coupling strength $1/\alpha$ broadens the MZM,
it has no effect on the height of zero-bias peak,
because whatever the effective coupling strength in $\bar{H}_T$ is,
it stands for the coupling term of both electron and hole leads,
so the e-h tunneling is always resonance at $V=0$,
which means a complete Andreev reflection and $G(0)=2e^2/h$.
The middle of the 0th e-h channel, for this reason, is perfect transparent.
In some experiments, the suppression of the height of the zero-bias peak happens.\cite{ZBPep9,ZBPep5}
Here we can exclude the influence of the phonons on the suppression.
The behavior of the zero-bias peak is very different with the
photon-assisted Majorana induced resonance Andreev reflection,
in which the height of the $0$th peak of the conductance is reduced by
the time-periodic potential.\cite{WP1}
The transport in non-zeroth channel is achieved by absorbing and emitting phonons
and taking the road of the 0th channel.
So the transparency in the middle of the $n$th channel is only related to EPI
but irrelevant to the normal lead-superconductor coupling strength $1/\alpha$.
The influence of EPI on transport property is presented in Fig.\ref{fig5}.
When EPI is absent, there is only a zero-bias peak in the tunneling spectroscopy $G(V)$.
With the increase of EPI strength $g$,
non-zeroth peaks come out at $V=n\omega_p$ [see Fig.\ref{fig5}(a)].
The width of peaks in $G(V)$ narrows with the increase of $g$ for the EPI-induced channel narrowing.
Meanwhile the heights of phonon sidebands $h_n=G(n\omega_p)$ rise with $g$,
which is illustrated in Fig.\ref{fig5}(b).
At zeros temperature and at the small EPI strength $g$,
the height of $n$th peak $h_n$ obeys the rule of
\begin{equation}
h_n\propto g^n.\label{Ghigh}
\end{equation}
This relation can be accounted for the following explanation:
In the condition of weak EPI,
the probability of $n$-phonon process is proportional to $\lambda^n$.
Take Fig.\ref{model}(b) as an example.
This diagram stands for single phonon process and contains one wavy line corresponding to $\lambda^1$.
The diagram represents the scattering coefficient $s_{eh}$,
and the conductance $G\propto|s_{eh}|^2$.
So the 1st conductance peak $h_1\propto\lambda^2\propto g^1$.
Similarly, the $n$th conductance peak $h_n\propto\lambda^{2n}\propto g^{n}$.
However, for the large EPI strength $g$,
the conductance peaks $h_n$ will deviate from this proportional relationship
($h_n\propto g^{n}$), because the conductance can not exceed $2e^2/h$.

\section{\label{sec4}conclusion}

In order to investigate the influence of electron-phonon interaction
on the Majorana induced resonance Andreev reflection,
a normal lead-superconductor junction is studied in this paper,
of which a single long wave phonon mode is coupled to the superconductor.
With the help of canonical transformation and mean-field approximation,
a non-equilibrium Green's function method is developed
for numerical research on this junction.
The Andreev reflection in the normal lead-superconductor junction
can be viewed as resonance tunneling from electron lead to hole lead,
where Majorana zero mode (MZM) plays the role of e-h channel.
With the phonon mode coupling, a series of subchannels at $E=n\omega_p$
are derived from the MZM channel with the same width but different transparency.
Electron in the subchannel can take the road of the MZM channel by emitting and absorbing phonons.
These channels are presented in LDOS $\rho(x,E)$ as a series of stripes.
The stripes are same in space distribution but different in scale.
These stripes are reduced to peaks in energy distribution $\rho(E)$ near
the normal lead-superconductor interface.
The lead-superconductor coupling strength broadens these peaks,
while electron-phonon interaction has two effects on these peaks.
One is the suppression in effective coupling strength
which narrows the width and raise the height.
The other one is the inter peak redistribution which changes the weights of every peak.
The tunneling spectroscopy presents conductance peaks similar to LDOS.
At zero temperature the width of peaks are the same in tunneling spectroscopy and LDOS.
At finite temperature, while the peaks in LDOS are slightly narrowed for the increase of phonon number,
the conductance peaks are severely broadened and strongly reduced for thermal broadening effect.
The height of the $n$th conductance peak is proportional to $g^n$ at zero temperature
and weak electron-phonon interaction strength $g$, presenting a feature of multi phonon process.

\section*{acknowledgments}
This work was financially supported by National Key R and D Program of China (Grant No. 2017YFA0303301),
NBRP of China (Grant No. 2015CB921102), NSF-China (Grant No. 11574007),
the Strategic Priority Research Program of Chinese Academy of Sciences (Grant No. XDB28000000),
and Beijing Municipal Science \& Technology Commission (Grant No.Z181100004218001).

\begin{appendix}

\section{spectral function}

In this Appendix A, we deduce the expression of the spectral function.
Using $\langle\langle\cdots\rangle\rangle$ and $\overline{\langle\langle\cdots\rangle\rangle}$
to denote the Green's function of $H$ and $\bar{H}$, respectively,
their relation is presented below:\cite{Mahan,FTF6,PART1,PAAR2}
\begin{eqnarray}
\langle\langle \hat{c}_x|\hat{c}^\dagger_x\rangle\rangle^<(\omega)=\sum_nL_n\overline{\langle\langle \hat{c}_x|\hat{c}^\dagger_x\rangle\rangle^<}(\omega+n\omega_p)\label{<},\\
\langle\langle \hat{c}_x|\hat{c}^\dagger_x\rangle\rangle^>(\omega)=\sum_nL_n\overline{\langle\langle \hat{c}_x|\hat{c}^\dagger_x\rangle\rangle^>}(\omega-n\omega_p)\label{>}.
\end{eqnarray}
The coefficient $L_n$ can be expressed by the $n$th modified Bessel function of first kind $I_n$:\cite{Mahan,FTF6,PART1,PAAR2}
\begin{equation}
L_n=\me^{-g(2N+1)}\me^{n\omega_p\beta/2}I_n(2g\sqrt{N(N+1)}),\label{Ln}
\end{equation}
with $\beta=1/k_B \mathcal{T}$.

The spectral function, i.e., the LDOS in the superconducting lead is defined as
\begin{equation}
\rho(x,\omega)=\frac{\mi}{2\piup}[\langle\langle \hat{c}_x|\hat{c}^\dagger_x\rangle\rangle^>(\omega)-\langle\langle \hat{c}_x|\hat{c}^\dagger_x\rangle\rangle^<(\omega)],\label{defA}
\end{equation}
which can be expressed by the transformed Green's function
\begin{eqnarray}
\rho(x,\omega)&=&\frac{\mi}{2\piup}\sum\limits_n L_n \left[\overline{\langle\langle \hat{c}_x|\hat{c}^\dagger_x\rangle\rangle}^>(\omega-n\omega_p) \right.\nonumber \\
&&\left.-\overline{\langle\langle \hat{c}_x|\hat{c}^\dagger_x\rangle\rangle}^<(\omega+n\omega_p)\right].\label{A0}
\end{eqnarray}
Hereafter the parameter $\omega$ is omitted if it does not lead to confusion.
Introduce the Nambu representation, and let\cite{SunAR3}
\begin{equation}
\bar{\mathcal{G}}(x,x')\equiv\begin{pmatrix}
\overline{\langle\langle \hat{c}_x|\hat{c}^\dagger_{x'}\rangle\rangle}&\overline{\langle\langle \hat{c}_x|\hat{c}_{x'}\rangle\rangle}\\
\overline{\langle\langle \hat{c}^\dagger_x|\hat{c}^\dagger_{x'}\rangle\rangle}&\overline{\langle\langle \hat{c}^\dagger_x|\hat{c}_{x'}\rangle\rangle}
\end{pmatrix}\label{Nam}
\end{equation}
to be the Green's function of $\bar{H}$ in Nambu representation.
With the help of Eq.(\ref{dis1}-\ref{dis3}),
we can directly achieve the retarded and advanced Green's function $\bar{\mathcal{G}}^{r/a}$.
The relations between the greater/lesser Green's function to the retarded/advanced one
are Keldysh equation\cite{NEGF1}
\begin{eqnarray}
\bar{\mathcal{G}}^<=\mi\bar{\mathcal{G}}^rf_L\bar{\Gamma}^L\bar{\mathcal{G}}^a\label{G<},\\
\bar{\mathcal{G}}^>=\bar{\mathcal{G}}^<-\mi\bar{\mathcal{A}}\label{G>},\\
\bar{\mathcal{A}}\equiv\mi[\bar{\mathcal{G}}^r-\bar{\mathcal{G}}^a].\label{A}
\end{eqnarray}
Here $f_L$ represents the Fermi distribution in the normal lead
\begin{equation}
f_L(\omega)=\begin{bmatrix}
f(\omega-\mu)& 0 \\
0& f(\omega+\mu),
\end{bmatrix}
\end{equation}
where $f(\omega)\equiv {1}/\{ \me^{{\omega}/{k_B \mathcal{T}}}+1\}$
and the chemical potential $\mu$ is controlled by bias voltage $V$ that
\begin{equation}
\mu\equiv eV/\hbar.
\end{equation}
The transformed linewidth function is defined as
\begin{equation}
\bar{\Gamma}^L=\begin{bmatrix}
\mi|\bar{T}_t|^2(\langle\langle \hat{d}_0|\hat{d}^\dagger_0\rangle\rangle_0^r-\langle\langle \hat{d}_0|\hat{d}^\dagger_0\rangle\rangle_0^a) & 0 \\
\hspace{-2cm}0 & \hspace{-2cm}\mi|\bar{T}_t|^2(\langle\langle \hat{d}^\dagger_0|\hat{d}_0\rangle\rangle_0^a-\langle\langle \hat{d}^\dagger_0|\hat{d}_0\rangle\rangle_0^r)\label{tgamma}
\end{bmatrix}
\end{equation}
where $\langle\langle\cdots\rangle\rangle_0$ stands for the Green's function of an isolated $H_L$.
Here $\bar{T}_t=T_t\me^{-g(N+1/2)}$ and $T_t=\frac{1}{\alpha}\frac{1}{ma_0^2}$
which is the tunneling term in Eq.(\ref{dis3}).
The surface Green's function $\langle\langle \hat{d}_0|\hat{d}^\dagger_0\rangle\rangle_0^{r/a}$
can be solved directly from Eq.(\ref{dis1}), so are $\bar{\Gamma}^L$.

Substituting Eqs.(\ref{G<},\ref{G>}) into Eq.(\ref{A0}), we have arrived at a numerical solvable expression of LDOS:
\begin{eqnarray}
\rho(\omega)&=&\frac{1}{2\piup}\sum\limits_{n}\left[L_n\bar{\mathcal{A}}(\omega_{-n})\right.\label{Aexp}\\
&+&\left.(L_{-n}-L_n)\mathcal{\bar{G}}^r(\omega_{-n})f_L(\omega_{-n})
\bar{\Gamma}^L(\omega_{-n})\mathcal{\bar{G}}^a(\omega_{-n})\right]_{11} \nonumber
\end{eqnarray}
where $\omega_n =\omega+n\omega_p$ and
the subscript ``11" represents the ``11" element of the matrix.

\section{\label{con}conductance}

In the Appendix B, we deduce the formulation for the conductance.
The electric current from the normal lead flowing into the superconductor
is the time derivative of the electron number
$N_L =\sum_{k} \hat{d}_k^{\dagger} \hat{d}_k$, that\cite{SunAR3,NEGF2}
\begin{equation}
I=-e\langle\dot{N}_L\rangle=-\frac{2e}{\hbar}\sum_{k,k'}\mathrm{Re}T_t\langle\langle \hat{d}_{k'}(t)|\hat{c}^\dagger_k(t)\rangle\rangle^<.\label{cur}
\end{equation}
With the help of the motion equation, the Green's function of interacting system $\langle\langle \hat{d}_{k'}|\hat{c}^\dagger_k\rangle\rangle^<(\omega)$ can be reduced to\cite{SunAR3,NEGF2}
\begin{eqnarray}
	\langle\langle \hat{d}_{k'}|\hat{c}^\dagger_k\rangle\rangle^<&=&\langle\langle \hat{d}_{k'}|\hat{d}^\dagger_{k'}\rangle\rangle_0^rT_t\sum_{k''}\langle\langle \hat{c}_{k''}|\hat{c}^\dagger_k\rangle\rangle^<\nonumber\\
	&+&\langle\langle \hat{d}_{k'}|\hat{d}^\dagger_{k'}\rangle\rangle_0^<T_t\sum_{k''}\langle\langle \hat{c}_{k''}|\hat{c}^\dagger_k\rangle\rangle^a.\label{motion}
\end{eqnarray}
Substituting Eq.(\ref{motion}) into Eq.(\ref{cur})
and transforming to the discrete Hamiltonian in the real space,
we arrive at
\begin{eqnarray}
I&=&-\frac{2e}{h}\int\dif\omega|T_t|^2\mathrm{Re}[\langle\langle \hat{d}_0|\hat{d}^\dagger_0\rangle\rangle_0^r\langle\langle \hat{c}_0|\hat{c}^\dagger_0\rangle\rangle^<\nonumber\\
&&+\langle\langle \hat{d}_0|\hat{d}^\dagger_0\rangle\rangle_0^<\langle\langle \hat{c}_0|\hat{c}^\dagger_0\rangle\rangle^a].\label{curr}
\end{eqnarray}
Here $\hat{d}_0 (\hat{d}_0^{\dagger})$ and $\hat{c}_0 (\hat{c}_0^{\dagger})$
are the annihilation (creation) operators at the end of the normal lead and superconductor,
respectively.
Introducing the linewidth function corresponding to Eq.(\ref{tgamma})
\begin{equation}
\Gamma^L=\begin{bmatrix}
\mi|T_t|^2(\langle\langle \hat{d}_0|\hat{d}^\dagger_0\rangle\rangle_0^r-\langle\langle \hat{d}_0|\hat{d}^\dagger_0\rangle\rangle_0^a) & 0 \\
\hspace{-2cm}0 & \hspace{-2cm}\mi|T_t|^2(\langle\langle \hat{d}^\dagger_0|\hat{d}_0\rangle\rangle_0^a-\langle\langle \hat{d}^\dagger_0|\hat{d}_0\rangle\rangle_0^r)
\end{bmatrix}
\end{equation}
and applying the fluctuation-dispassion theorem on $\langle\langle \hat{d}_0|\hat{d}^\dagger_0\rangle\rangle_0^<$, combining with
\begin{equation}
\langle\langle \hat{c}_0|\hat{c}^\dagger_0\rangle\rangle^r-\langle\langle \hat{c}_0|\hat{c}^\dagger_0\rangle\rangle^a=\langle\langle \hat{c}_0|\hat{c}^\dagger_0\rangle\rangle^>-\langle\langle \hat{c}_0|\hat{c}^\dagger_0\rangle\rangle^<,
\end{equation}
and Eqs.(\ref{<},\ref{>},\ref{Nam}), the Eq.(\ref{curr}) is reduced to
\begin{eqnarray}
I&=&\frac{\mi e}{h}\sum_nL_n\int\{\Gamma^L[(1-f_L(\omega))\bar{\mathcal{G}}^<(\omega_n)\nonumber\\
&&+f_L(\omega)\bar{\mathcal{G}}^>(\omega_{-n})]\}_{11}\dif\omega.
\end{eqnarray}
Substituting Eqs.(\ref{G<},\ref{G>},\ref{A}), we get a solvable expression of electric current:
\begin{widetext}
\begin{eqnarray}
I&=&\frac{e}{h}\int\dif\omega\sum_n
 \left\{L_{-n}\Gamma^L(\omega)f_L(\omega)\bar{\mathcal{A}}(\omega_n)
 -\Gamma^L(\omega)[L_n+(L_{-n}-L_n)f_L(\omega)]
  \mathcal{\bar{G}}^r(\omega_n)f_L(\omega_n)\bar{\Gamma}^L(\omega_n)
  \mathcal{\bar{G}}^a(\omega_n)\right\}_{11}\nonumber\\
&=&\frac{e}{h}\int\dif\omega\sum_n \left\{L_{-n}\Gamma^L_{11}(\omega)f(\omega-\mu)
 \bar{\mathcal{A}}_{11}(\omega_n) \right.\nonumber\\
&-&\left.\Gamma^L_{11}(\omega)[L_n+(L_{-n}-L_n)f(\omega-\mu)]
\left[\mathcal{\bar{G}}_{11}^r(\omega_n)f(\omega_n-\mu)\bar{\Gamma}^L_{11}(\omega_n)
 \mathcal{\bar{G}}_{11}^a(\omega_n)
+\mathcal{\bar{G}}_{12}^r(\omega_n)f(\omega_n+\mu)\bar{\Gamma}^L_{22}(\omega_n)
 \mathcal{\bar{G}}_{21}^a(\omega_n)\right]\right\} . \label{fincur}
\end{eqnarray}
Then, the differential conductance $G(V) \equiv {\dif I}/{\dif V} $ is
\begin{eqnarray}
 G(V)&=&
 \frac{e^2}{h}\frac{1}{T}\int\dif\omega\sum_n
 \left\{L_{-n}\Gamma^L(\omega)\sigma_z
  f_L(\omega)(1-f_L(\omega))\bar{\mathcal{A}}(\omega_n)\right.\nonumber\\
&-&\Gamma^L(\omega)[L_n+(L_{-n}-L_n)f_L(\omega)]\mathcal{\bar{G}}^r(\omega_n)\sigma_zf_L(\omega_n)
(1-f_L(\omega_n))\bar{\Gamma}^L(\omega_n)\mathcal{\bar{G}}^a(\omega_n)\\
&-&\left.\Gamma^L(\omega)(L_{-n}-L_n)\sigma_zf_L(\omega)(1-f_L(\omega))
 \mathcal{\bar{G}}^r(\omega_n)f_L(\omega_n)\bar{\Gamma}^L(\omega_n)\mathcal{\bar{G}}^a(\omega_n)
\right\}_{11}.\nonumber
\end{eqnarray}
In particular, at zero temperature the differential conductance $G(V)$ reduces into:
\begin{eqnarray}
G(V) &=&
 \frac{e^2}{h}\left\{\sum_nL_{-n}\Gamma^L_{11}(\mu)\bar{\mathcal{A}}_{11}(\mu_n)
-\sum_n\Gamma^L_{11}(\mu_{-n})L_n\mathcal{\bar{G}}_{11}^r(\mu)\bar{\Gamma}^L_{11}(\mu)\mathcal{\bar{G}}_{11}^a(\mu)
\right.\nonumber\\
&&-\sum_{n\leq0}\Gamma^L_{11}(\mu)(L_{-n}-L_n)\mathcal{\bar{G}}_{11}^r(\mu_n)\bar{\Gamma}^L_{11}(\mu_n)
  \mathcal{\bar{G}}_{11}^a(\mu_n)
-\sum_{n>0}\Gamma^L_{11}(\mu_{-n})(L_{-n}-L_n)\mathcal{\bar{G}}_{11}^r(\mu)\bar{\Gamma}^L_{11}(\mu)\mathcal{\bar{G}}_{11}^a(\mu)\nonumber\\
&&+\sum_n\Gamma^L_{11}(-\mu_n)L_n\mathcal{\bar{G}}_{12}^r(-\mu)\bar{\Gamma}^L_{22}(-\mu)\mathcal{\bar{G}}_{21}^a(-\mu)
-\sum_{n<n_0}\Gamma^L_{11}(\mu)(L_{-n}-L_n)\mathcal{\bar{G}}_{12}^r(\mu_n)\bar{\Gamma}^L_{22}(\mu_n)
 \mathcal{\bar{G}}_{21}^a(\mu_n)\nonumber\\
&&+\left.\sum_{n>n_0}\Gamma^L_{11}(-\mu_n)(L_{-n}-L_n)\mathcal{\bar{G}}_{12}^r(-\mu)\bar{\Gamma}^L_{22}(-\mu)\mathcal{\bar{G}}_{21}^a(-\mu)
\right\},
\end{eqnarray}
where $\mu_n=\mu+n\omega_p$ and $n_0=-\frac{2\mu}{\omega_p}$.
\end{widetext}

\end{appendix}

\end{document}